**Title:** Electronic Health Record Phenotyping with Internally Assessable Performance (PhIAP) using Anchor-Positive and Unlabeled Patients

**Authors:**

Lingjiao Zhang, MS[1], Xiruo Ding, MS[2], Yanyuan Ma, PhD[3], Naveen Muthu, MD[4], Imran Ajmal, MD[2], Jason H. Moore, PhD[1], Daniel S. Herman, MD, PhD[2*], Jinbo Chen, PhD[1*]

[1]Department of Biostatistics, Epidemiology and Informatics, Perelman School of Medicine, University of Pennsylvania, Philadelphia, Pennsylvania, USA

[2]Department of Pathology and Laboratory Medicine, University of Pennsylvania, Philadelphia, Pennsylvania, USA

[3]Department of Statistics, Penn State University, Philadelphia, Pennsylvania, USA

[4]Department of Biomedical and Health Informatics, University of Pennsylvania, Philadelphia, Pennsylvania, USA

**\*Corresponding authors:**
Jinbo Chen
203 Blockley Hall
423 Guardian Drive
Philadelphia, PA 19104
Office: 215-746-3915
Fax: 215-573-1050
Email: jinboche@pennmedicine.upenn.edu

Daniel Herman
3400 Spruce Street
Hospital of the University of Pennsylvania
Founders 7.103
Philadelphia, PA 19104
Office: 215-662-6575
Fax: 215-662-7529
Email: Daniel.herman2@uphs.upenn.edu

**Keywords:** Electronic health record [D057286], clinical decision support systems [D020000], biostatistics [D056808], statistical models [D015233], hypertension [D006973]

**Word count:** 3941

**ABSTRACT**

**Objective**: Building phenotype models using electronic health record (EHR) data conventionally requires manually labeled cases and controls. Assigning labels is labor intensive and, for some phenotypes, identifying gold-standard controls is prohibitive. To facilitate comprehensive clinical decision support and research, we sought to develop an accurate EHR phenotyping approach that assesses its performance without a validation set.

**Materials and Methods**: Our framework relies on specifying a random subset of cases, potentially using an anchor variable that has excellent positive predictive value and sensitivity that is independent of predictors. We developed a novel maximum likelihood approach that efficiently leverages data from anchor-positive and unlabeled patients to develop logistic regression phenotyping models. Additionally, we described novel statistical methods for estimating phenotyping prevalence and assessing model calibration and predictive performance measures.

**Results**: Theoretical and simulation studies indicated our method generates accurate predicted probabilities, leading to excellent discrimination and calibration, and consistent estimates of phenotype prevalence and anchor sensitivity. The method appeared robust to minor lack-of-fit and the proposed calibration assessment detected major lack-of-fit. We applied our method to EHR data to develop a preliminary model for identifying patients with primary aldosteronism, which achieved an AUC of 0.99 and PPV of 0.8.

**Discussion**: We developed novel statistical methods for accurate model development and validation with minimal manual labeling, facilitating development of scalable, transferable, semi-automated case labeling and practice-specific models.

**Conclusion**: Our EHR phenotyping approach decreases labor-intensive manual phenotyping and annotation, which should enable broader model development and dissemination for EHR clinical decision support and research.

## BACKGROUND AND SIGNIFICANCE

The adoption of Electronic Health Records (EHRs) by healthcare systems has the potential to enable implementation of comprehensive, computational clinical decision support and clinical research[1-4]. However, EHRs have been designed primarily to support documentation for medical billing rather than being intricately embedded in clinical diagnostic processes[5-7], so patients' complex, clinical phenotypes are not natively represented in the rich, accurate, precise format that is important for many analyses[8-11]. To overcome this limitation, a variety of heuristic and statistical methods have been developed for phenotyping patients based on EHR data[12,13]. The vast majority of these existing methods require a curated dataset of patients who are completely and accurately labeled with regard to the presence or absence of a phenotype. Such methods require experts to retrospectively review EHR charts and/or prospectively evaluate patients to label them with respect to the phenotype. For many phenotypes, the labor and cost of these processes limit the achievable sample size, compromising the accuracy of potential phenotyping methods.

Such approaches can be improved upon by appreciating that clinical practice workflows are often not symmetric with regards to cases and controls. In practice, for most diseases, only specific patients are actively evaluated based on clinical suspicion. However, the converse is not true; there are very few phenotypes for which everyone is actively screened for, so clinical practice data is frequently insufficient to identify a set of gold-standard controls. As a result, for many phenotypes, the most readily accessible annotations are an incomplete set of gold-standard cases and few or no gold-standard controls. It is highly desirable to develop new phenotyping methods that efficiently and accurately leverage such incomplete phenotyping, or "positive-only" data.

Within this framework, one special case with extremely desirable consequences is when the labeled cases are a random subset of the full set of true cases[14]. This set of cases could be identified through an active labeling process as part of existing clinical care or research. However, a more generalizable strategy that has the potential to decrease the requirement for manual chart review is the use of binary "anchor variables"[15] that summarize clinical domain expertise for classifying patients' phenotype. By definition, an anchor variable has perfect positive predictive value (PPV), but is not required to have high sensitivity. That is, anchor positivity indicates presence of the phenotype, but anchor negativity is non-deterministic of the true phenotype status. The second critically important requirement for an anchor variable is that its sensitivity be independent of all phenotype model predictors. An ideal anchor variable is a structured data element in the EHR that is only present in cases, such as the result of a diagnostic confirmatory test or an order that only follows a definitive diagnosis. For example, a pathologic diagnosis of cancer will in most scenarios have a very high PPV, but perhaps an imperfect sensitivity because of variability in practice or documentation, variability in diagnostic categories, or data incompleteness. For many phenotypes, such definitive diagnostic information may not be available, so surrogates such as diagnosis codes, medications, or note concepts must be considered. Expert knowledge is necessary to select a variable or a composite variable that meets the high PPV and predictor independence requirements[15].

A method for learning from such positive-only data was proposed by Elkan and Noto[14]. Their method predicted the probability of phenotype presence through estimating the probability that a subject is anchor positive, motivated by a key observation that the two probabilities differ only by a constant factor, the anchor sensitivity. This method requires establishment of the true phenotype status for a random subset of patients to obtain a consistent estimate of the anchor sensitivity and phenotype prevalence. Unfortunately, when phenotype prevalence is low, the advantage of positive-only design is lost because a large validation subsample is needed for accurate estimation of anchor sensitivity. In addition, while this method yields a consistent estimate of phenotype prevalence when the predictor distributions for cases and controls are completely separable[16], the estimated phenotype probability may fall outside of the [0,1] range if the predictor distributions overlap, compromising the calibration of the resultant prediction model. Another related approach to semi-automated learning of EHR phenotypes assumes random label errors and is most suitable for phenotypes that are not rare. This method does not require a validation sample for modeling, but does to evaluate method performance[17-19].

One application for which a positive-only design is advantageous is identifying patients and estimating the prevalence of primary aldosteronism (PA) in a clinical practice. Based on epidemiological studies, PA is the most common cause of secondary hypertension, thought to affect ~5% of hypertensive patients and up to 20% of patients with resistant hypertension[20-22]. PA can be treated effectively by unilateral adrenalectomy or targeted medications. Previous methodological work has tried to improve the diagnostic evaluation for PA amongst evaluated patients[24-26]. Unfortunately, PA is not recognized in many affected patients[23], so methods for EHR phenotyping have the potential to dramatically improve care for these patients.

To enable more efficient and accurate EHR phenotyping in the setting of incomplete clinical training phenotypes, we propose a maximum likelihood method to develop a logistic regression prediction model using positive-only EHR data that does not require an external validation sample for parameter estimation. We have demonstrated, via extensive simulation and application to identify PA patients in our institution's EHR, that this method develops models that accurately identify unlabeled cases and yields consistent estimates of phenotype prevalence, anchor sensitivity, and model performance.

## MATERIALS AND METHODS

*Positive-only data*

Let $Y$ denote the binary label for the phenotype (1: case; 0: control), $X$ denote a vector of clinical variables that are predictive of $Y$, and $S$ denote a binary indicator denoting the anchor variable being positive (1: positive; 0: negative). Here $(X, Y, S)$ are considered as random variables with joint distribution $p(X, Y, S)$ from which EHR patients, including both anchor-positive cases and unlabeled patients, are randomly drawn. For a well-chosen anchor variable, its high-PPV property can be stated formally as $p(Y = 1|S = 1) = 1$. But given its imperfect sensitivity, $Y$ can take either value 1 or 0 when $S = 0$. Let $N$ denote the total number of patients. Only $(\boldsymbol{X}, S)$ are observed for all $N$ patients. We use a logistic regression working model to relate $Y$ and $\boldsymbol{X}$ which is commonly implemented in the EHR setting, although our method is applicable for any parametric model that is reasonable for modeling binary outcome variables:

$$\text{logit } p(Y = 1|\boldsymbol{X}; \beta) = \boldsymbol{X}^T \beta. \tag{1}$$

Here we allow $\boldsymbol{X}$ to include a vector of 1 so that the intercept parameter is implicitly included in the logit function $\boldsymbol{X}^T \beta$. For notational simplicity, we use $P(\boldsymbol{X}; \beta)$ to denote $p(Y = 1|\boldsymbol{X}; \beta)$. Below, we present the likelihood method when anchor sensitivity is a constant.

*Phenotyping with an anchor of constant sensitivity*

The constant anchor sensitivity is formalized as conditional independence[15],

$$p(S = 1|Y = 1, \boldsymbol{X}) = p(S = 1|Y = 1) \equiv c, \tag{2}$$

where $c$ is a constant between 0 and 1. For the population defined by $p(\boldsymbol{X}, Y, S)$, let $F(\boldsymbol{X})$ denote the cumulative distribution function of $\boldsymbol{X}$, $f(\boldsymbol{X})$ the corresponding probability density function, and $q$ the phenotype prevalence, i.e. $q = p(Y = 1) = \int P(\boldsymbol{X}; \beta) dF(\boldsymbol{X})$. Let $h$ be the probability of anchor being positive, $h = p(S = 1)$. It is easy to show that anchor sensitivity $c$ equals $h/q$ by applying (2). We propose to estimate odds ratio parameters $\beta$ and anchor sensitivity $c$ through maximizing the likelihood function, which is the probability of the observed data for the $N$ patients:

$$L(\beta, c) = \prod_{i=1}^{N} p(\boldsymbol{X}_i, S_i = 1)^{S_i} p(\boldsymbol{X}_i, S_i = 0)^{1-S_i}$$

$$\propto \prod_{i=1}^{N} \{cp(\boldsymbol{X}_i; \beta)\}^{S_i} \{1 - cp(\boldsymbol{X}_i; \beta)\}^{1-S_i}$$

As shown in Appendix A, $(\beta, c)$ are identifiable with positive-only data. The resultant estimates $(\hat{\beta}, \hat{c})$ can be obtained by maximizing the log likelihood function $\log L(\beta, c)$. The large sample variance-covariance matrix of these estimates can be obtained from the inverse of the information matrix. We propose two methods for estimating phenotype prevalence $q$. Because $q$ can be expressed as $h/c$, it can be estimated as $\hat{q} = \hat{h}/\hat{c}$, where $\hat{h}$ is the maximum likelihood estimate of $p(S = 1)$ and equal to the sample fraction of those with $S = 1$. Alternatively, it can be estimated as the average of the estimated phenotype probabilities, $N^{-1}\sum_{i=1}^{N} P(\mathbf{X}_i, \hat{\beta})$.

*Model calibration and predictive accuracy measures*

It is important to validate the performance of model (1) for predicting $Y$ with respect to calibration and predictive accuracy. But this is not straightforward with positive-only data because $Y$ is unobserved for the unlabeled. For calibration, Hosmer-Lemeshow type of goodness-of-fit assessment requires a subsample of the unlabeled to have validated $Y$ status. Here we propose an intermediate approach. Because anchor-positive cases have definitive phenotype status, we focus on model calibration among the unlabeled. Since the true status of $Y$ is unobserved, we propose to compare a non-parametric estimate of case numbers with model-predicted case numbers in intervals defined by a priori cutoff points for $P(\mathbf{X}_i, \hat{\beta})$. Within interval $a < P(\mathbf{X}_i, \hat{\beta}) < b$, where $a$ and $b$ are two positive numbers in the range of (0,1), we can show that

$$p\{Y = 1 | a < P(\mathbf{X}_i, \hat{\beta}) < b, S = 0\} = \frac{p\{a < P(\mathbf{X}_i, \hat{\beta}) < b | Y = 1, S = 0\} p\{Y = 1, S = 0\}}{p\{a < P(\mathbf{X}_i, \hat{\beta}) < b, S = 0\}}$$
$$= \frac{(q - h) p\{a < P(\mathbf{X}_i, \hat{\beta}) < b, S = 1\}}{h p\{a < P(\mathbf{X}_i, \hat{\beta}) < b, S = 0\}}.$$

A nonparametric estimate of $p\{Y = 1 | a < P(\mathbf{X}_i, \hat{\beta}) < b, S = 0\}$ can then be obtained as

$$\hat{p}^n\{Y = 1 | a < P(\mathbf{X}_i, \hat{\beta}) < b, S = 0\} = \frac{(q^* - N^{-1}\sum_{i=1}^{N} S_i) N^{-1} \sum_{i=1}^{N} I\{a < P(\mathbf{X}_i, \hat{\beta}) < b, S_i = 1\}}{(N^{-1}\sum_{i=1}^{N} S_i) N^{-1} \sum_{i=1}^{N} I\{a < P(\mathbf{X}_i, \hat{\beta}) < b, S_i = 0\}}.$$

Here $q^*$ can be the MLE derived above, $\hat{q}$, which makes the estimate $\hat{p}^n$ not completely "nonparametric". But it can also be an educated guess by clinicians or from other investigations, which is often available for EHRs. Let $\hat{p}^m\{Y = 1 | a < P(\mathbf{X}_i, \hat{\beta}) < b, S = 0\}$ denote the model-based estimate of $p\{Y = 1 | a < P(\mathbf{X}_i, \hat{\beta}) < b, S = 0\}$. We have

$$\hat{p}^m\{Y = 1 | a < P(\mathbf{X}_i, \hat{\beta}) < b, S = 0\} = \frac{\sum_{i=1}^{N}(1 - \hat{c}) P(\mathbf{X}_i, \hat{\beta}) I\{a < P(\mathbf{X}_i, \hat{\beta}) < b\}}{\sum_{i=1}^{N}\{1 - \hat{c} P(\mathbf{X}_i, \hat{\beta})\} I\{a < P(\mathbf{X}_i, \hat{\beta}) < b\}}.$$

We propose to compare $\hat{p}^n$ and $\hat{p}^m$ for assessing model calibration, where similar values of $\hat{p}^n\{Y = 1 | a < P(X_i, \hat{\beta}) < b, S = 0\}$ and $\hat{p}^m\{Y = 1 | a < P(X_i, \hat{\beta}) < b, S = 0\}$ indicate good calibration.

For a well-calibrated model $P(X; \beta)$ as indicated by our method above, we study the predictive accuracy of the model among the unlabeled. To evaluate the accuracy of model (1) for classification, we consider statistical measures including true positive rate ($TPR_v$), false positive rate ($FPR_v$), positive predictive value ($PPV_v$), negative predictive value ($NPV_v$) at a decision threshold $v$ and area under the $ROC$ curve ($AUC$). Given that anchor-positive patients are classified as cases by definition, we propose to assess model performance only in the unlabeled patients. Interestingly, we found that these measures can all be estimated using positive-only data without requiring that model (1) reflects the true relationship between $Y$ and $X$, a feature that is particularly attractive for EHR data. $TPR_V$ is defined as

$$TPR_v = p\{P(X; \beta) > v | Y = 1, S = 0\}.$$

The constant anchor sensitivity requirement warrants that $S$ and $X$ are independent given $Y = 1$. Therefore, $TPR_v$ for the unlabeled is equal to that for anchor-positive cases, $TPR_v = p\{P(X; \beta) > v | Y = 1, S = 1\}$. This implies that $TPR_v$ for the unlabeled can be estimated using anchor-positive cases as

$$\widehat{TPR}_v = \hat{p}\{P(X; \beta) > v | S = 1\} = \frac{N^{-1} \sum_{i=1}^{N} I\{P(X_i; \hat{\beta}) > v, S_i = 1\}}{\hat{h}}.$$

The $FPR_v$ is defined and further-rewritten as

$$FPR_v = p\{P(X; \beta) > v | Y = 0, S = 0\} = \frac{p\{P(X; \beta) > v, S = 0\} - p\{P(X; \beta) > v | S = 1\} p(Y = 1, S = 0)}{p(Y = 0)}.$$

Therefore, it can be estimated similarly as

$$\widehat{FPR}_v = \frac{N^{-1} \sum_{i=1}^{N} I\{P(X_i; \hat{\beta}) > v, S_i = 0\}}{(1 - \hat{q})} - \frac{(\hat{q} - \hat{h}) N^{-1} \sum_{i=1}^{N} I\{P(X_i; \hat{\beta}) > v, S_i = 1\}}{\hat{h}(1 - \hat{q})},$$

Where $\hat{h} = N^{-1} \sum_{i=1}^{N} S_i$, $\hat{q} = N^{-1} \sum_{i=1}^{N} P(X_i, \hat{\beta})$.

Similar estimates of $PPV_v$, $NPV_v$, and $AUC$ are provided in Appendix C. We obtain standard errors of the estimated predictive accuracy measures based on 1000 bootstrap iterations.

*Simulation studies*

We carried out extensive simulation studies to evaluate the performance of our method (See Supplementary Material, Appendix D).

*Clinical data*

Demographics, laboratory results, encounter metadata, diagnosis codes, and unstructured clinical text were retrospectively queried and extracted from between January 1, 1997 and July 1, 2017 from the University of Pennsylvania Health System (UPHS) clinical data repository and PennChart EPIC Clarity reporting database. Individual laboratory test results and vital signs were selected that were temporally closest to, and no more than 14 days from, the date of the blood aldosterone testing. Diagnosis codes for primary aldosteronism or other hyperaldosteronism were counted. To describe patients' time in our healthcare system, we calculated pairwise time intervals between each patient's first encounter, last encounter, initial PA screening laboratory testing, and PA evaluative adrenal vein sampling (AVS) procedure. Notes were queried using a series of regular expressions (Table S1) and counts were summed per patient. Variables with highly positively skewed distributions were log-transformed, and all continuous variables were standardized. Code and full data dictionaries are available in *http://bitbucket.com/hermanlab/PhIAP*.

Analyses were restricted to the subset of patients with complete data on included variables, simplistically assuming that the missingness (Table S2) was completely at random. PA anchor variable was implemented initially leveraging an existing PA research database curated through 2015[27] (Anchor 1) and later supplemented by adding positive cases that had laboratory results for an AVS procedure (Anchor 2). Patient charts for all patients identified by the two final positive-only models with $\hat{p}(Y = 1|X) \geq 0.2$, 2+ diagnosis codes, or strict laboratory criteria (blood aldosterone $\geq$ 15 ng/Dl, renin activity $\leq$ 0.5 ng/mL/hr, aldosterone:renin $\geq$ 30) were reviewed by clinician I.A. and unclear cases were further reviewed by D.S.H. This study was reviewed and approved by the University of Pennsylvania IRB.

## RESULTS

**Simulation studies**

Our proposed method generated consistent and efficient estimates for $\beta$, $c$, $q$, and predictive accuracy measures across a wide range of phenotype prevalence (Table S3-S6) and anchor sensitivity (Tables S7, S8). Model lack-of-fit, deviation of the fitted model from the data generating model, could impact the parameter estimation and predictive accuracy. Simulation showed that the parameter estimators were relatively robust to minor model misspecifications, but did deviate with major model misspecifications (Tables S4). To address this, we showed that our proposed method of assessing calibration can detect such deviations (Table S5). In addition, since anchor sensitivity is not always fixed, we showed that in the setting of known, multi-level anchor sensitivity, stratified modeling yielded better performance (Table S10-S12).

**Application to identify patients with diagnosed primary aldosteronism in EHR data**

To identify PA patients in our healthcare system, we extracted EHR data for 6,319 patients who had an order for a PA screening laboratory test, blood aldosterone concentration, between January 01, 1997 and July 01, 2017. We first selected as training cases patients with definitive PA based on a documented history of an AVS procedure, for PA evaluation, or unilateral adrenalectomy, for PA management. We considered two derivations of this composite anchor phenotype, which we will refer to as anchor 1 and anchor 2. Anchor 1 labeled 101 (2.8%) patients as PA cases and the anchor 2 labeled an additional 34 patients, increasing the set of PA patients to a total of 135 (3.8%). Next, to identify additional PA patients among the unlabeled patients, we considered two rule-based algorithms: presence of 2 or more diagnosis codes or the presence of PA laboratory testing results that meet conservative diagnostic criteria. Based on chart review, these heuristics demonstrated excellent PPVs of 85% and 90%, respectively. However, they appeared to have only modest sensitivity, because they failed to flag all anchor-positive patients.

To identify further PA patients, we trained models using our novel method EHR phenotyping with internally assessable performance (PhIAP) using anchor-positive and unlabeled patients. We first separately developed preliminary logistic regression models trained using each of the anchor phenotypes. We selected and engineered potential predictors based on clinical expertise and univariate logistic regression models for predicting anchor positivity, ultimately including four demographic variables, six variables for laboratory results, eight variables for summarizing encounter metadata, two variables aggregating diagnosis codes, and six variables extracted from unstructured clinical text (Table S2). We applied our proposed method, starting with a baseline model that included only variables available at the time of PA diagnostic laboratory screening, including demographics, blood pressure, and time in health system prior to PA screening (Tables S13, S14). Then we serially added further sets of variables until all candidate predictors were included. Finally, we performed backward stepwise variable selection until all included variables had *p*-values less than 0.1. Upon determination of a final subset of predictors, we fitted the final prediction model among the 3579 patients who had complete data for these included variables (Tables S15, S16). This final population had demographics and laboratory results comparable to that of the full cohort (Table S17).

The baseline model using anchor 1 estimated the phenotype prevalence $\hat{q}$ as 0.30 (estimated asymptotic standard error (SE): 0.05) and anchor sensitivity $\hat{c}$ as 0.10 (SE: 0.02). This $\hat{q}$ was much higher than expected. As the subsequent four sets of variables were sequentially added, $\hat{c}$ became 0.58, 0.64, 0.49 and 0.45, respectively. The corresponding prevalence estimates for $\hat{q}$ were 0.05, 0.05, 0.07 and 0.08 (Table S13). Results were similar when anchor 2 was used to label cases, except that the estimated anchor sensitivities were higher (Table S14). For both anchors, $\hat{c}$ and $\hat{q}$ estimates from the baseline model appeared very different from the subsequent, richer models. To probe the baseline model, we applied our proposed method for evaluating calibration (Table S18). The nonparametrically-estimated and model-predicted case numbers in the pre-defined intervals of predicted probabilities were quite different, suggesting poor calibration. For example, the nonparametrically-estimated case number was 69 in interval $0.7 < P(X_i, \hat{\beta}) < 0.8$, compared to model-predicted case number of 43.

For the final model using anchor 1, $\hat{q}$ was estimated to be 0.08 (SE: 0.01) and $\hat{c}$ was estimated to be 0.37 (SE: 0.05), which predicted that there were 171 unlabeled PA patients. With Anchor 2, $\hat{c}$ increased to 0.55 (SE: 0.04), which was expected because Anchor 2 labeled more PA patients. The PA prevalence was estimated to be 0.07 (SE: 0.01). For both anchors, the final prediction model appeared well calibrated, with similar nonparametrically-estimated and model-predicted case numbers in each fixed interval of predicted probabilities (Table 1). The final prediction model also appeared to have high discriminatory power. The estimated probabilities of PA presence, $\hat{p}(Y = 1|X)$ showed good separation between groups, with anchor-positive cases mostly having high predicted probabilities (Figure 1).

**Table 1.** Model-predicted vs. nonparametrically-estimated case numbers in pre-defined intervals of predicted probabilities. The two estimated numbers were compared in each interval of predicted probabilities to inform model calibration.

| Interval | Anchor 1 | | Anchor 2 | |
| --- | --- | --- | --- | --- |
| | Model-predicted | Nonparametrically-estimated | Model-predicted | Nonparametrically-estimated |
| 0.0_0.1 | 11 | 12 | 7 | 7 |
| 0.1_0.2 | 6 | 8 | 4 | 4 |
| 0.2_0.3 | 6 | 3 | 3 | 6 |
| 0.3_0.4 | 5 | 7 | 3 | 1 |
| 0.4_0.5 | 6 | 7 | 3 | 5 |
| 0.5_0.6 | 9 | 8 | 5 | 3 |
| 0.6_0.7 | 5 | 0 | 3 | 3 |
| 0.7_0.8 | 10 | 8 | 5 | 5 |
| 0.8_0.9 | 13 | 10 | 7 | 6 |
| 0.9_1.0 | 98 | 99 | 82 | 76 |

We evaluated predictive accuracy of the final selected model using 10-fold cross validation and reported empirical standard errors (ESE) based on 1000 bootstrap iterations. For both anchors, the model achieved an averaged AUC of 0.99 (ESE: 0.005). With anchor 1, the respective estimates of PPV and TPR were 0.65 (ESE: 0.10) and 0.89 (ESE: 0.03) at decision threshold 0.2, and 0.83 (ESE: 0.05) and 0.79 (ESE: 0.06) at threshold 0.5. With anchor 2, these estimates were 0.63 (ESE: 0.11) and 0.91 (ESE: 0.04) at threshold 0.2 and 0.82 (ESE: 0.08) and 0.82 (ESE: 0.06) at threshold 0.5 (Table 2). Chart review of the patients the models predicted to be diagnosed with PA revealed similar PPVs of 74% and 80% (anchor 1) and 74% and

77% (anchor 2) at decision thresholds of 0.2 and 0.5, respectively. Notably, among PA patients not flagged by either of the originally evaluated rule-based algorithms, these models identified 7 additional PA patients.

**Table 2.** Estimated predictive accuracy measures (Empirical Standard Error) of resultant classifiers with each of the two anchors. The AUC was estimated to be 0.993 (ESE: 0.005) for both anchors.

|        | Anchor1 |         |         |         | Anchor 2 |         |         |         |
|--------|---------|---------|---------|---------|----------|---------|---------|---------|
| Cutoff | 0.2     | 0.3     | 0.4     | 0.5     | 0.2      | 0.3     | 0.4     | 0.5     |
| PPV    | 0.653   | 0.736   | 0.791   | 0.833   | 0.629    | 0.709   | 0.765   | 0.822   |
|        | (0.101) | (0.079) | (0.062) | (0.047) | (0.112)  | (0.104) | (0.093) | (0.080) |
| TPR    | 0.894   | 0.865   | 0.837   | 0.786   | 0.906    | 0.876   | 0.847   | 0.816   |
|        | (0.031) | (0.039) | (0.047) | (0.056) | (0.035)  | (0.044) | (0.053) | (0.062) |
| NPV    | 0.994   | 0.993   | 0.992   | 0.989   | 0.997    | 0.996   | 0.995   | 0.994   |
|        | (0.001) | (0.002) | (0.002) | (0.002) | (0.001)  | (0.001) | (0.002) | (0.002) |
| FPR    | 0.024   | 0.016   | 0.011   | 0.008   | 0.017    | 0.012   | 0.008   | 0.006   |
|        | (0.005) | (0.004) | (0.002) | (0.002) | (0.005)  | (0.003) | (0.002) | (0.002) |

**DISCUSSION**

A key step to comprehensive and accurate EHR phenotyping is the development of accurate algorithms that efficiently leverage clinical expertise and available data. Currently, most existing methods require a complete set of expert-annotated cases and controls to train phenotyping algorithms or incorrectly treat unlabeled patients as definitive controls. Our proposed method builds accurate model classifiers based upon a random sample of positive cases and a large number of unlabeled patients. To identify a random sample of cases, the method can leverage domain expertise summarized in the form of an anchor variable, with modest upfront effort from clinical experts[15]. Compared to standard strategies, this method dramatically decreases the need for labor-intensive chart annotation and prospective phenotyping.

The model performance relies upon the independence of anchor sensitivity and model predictors, which necessitates meticulous selection of the anchor based on clinical expertise. If unclear, the appropriateness of an anchor could be supported by explicitly validating the estimated phenotype prevalence, model sensitivity, or the conditional independence assumption. We refer readers to Halpern's work[15,28] for methods to support defining potential anchors. In this work, we also extended the current implementation of anchor variables to allow anchor sensitivity to vary across a fixed number of discrete strata that are defined by patient EHR data (Appendix B). Recent work has considered phenotyping methods that take advantage of both noisy labels with random error and anchor variable framework[18]. Our method can be extended in this regard.

We proposed novel statistical methods for assessing calibration and predictive accuracy of the developed prediction model. A model must be well-calibrated for predictive accuracy measures to meaningfully inform model performance. Classically, validating models requires annotated controls or labels for a random sample of patients. To assess calibration, we developed a novel approach that took advantage of the defining characteristics of an anchor variable to use only the observed data for anchor-labeled cases and the unlabeled patients. In the analysis of Penn EHR data and our simulation studies, we showed that poor calibration of the model or breaking the conditional independence assumption can lead to severely biased estimates of anchor sensitivity and phenotype prevalence. Fortunately, our proposed calibration method can help detect poor calibration, and the estimates appeared to be largely consistent as long as our method indicates good calibration even in the setting of minor model misspecification. This is very important, because a good estimate of anchor sensitivity and its confidence interval can reveal the extent of undetected cases among the unlabeled patients and identify individual unlabeled patients. It is challenging to confidently scale previous methods for positive-only model learning to large numbers of phenotypes, because of the delicacy of the conditional independence assumption and the necessity for external validation. However, PhIAP's internal assessment of model performance should enable it to scale robustly.

Another attractive feature of this method is transferability to other practices. The anchor concept may itself be more easily transferred rather than the full model[18]. At secondary sites, chart review need only be performed to confirm that the anchor has very high PPV for the single phenotype of interest, which is considerably less burdensome validation of a classically fit or transferred model. Unlike previous implementations of such positive-only model development, because of the internal assessment of model performance, if a model can be fit locally with good performance parameters, then there is no absolute need for additional chart review.

In this work, we applied our method to develop models to identify patients with PA. In selecting as anchor variables the PA subtyping diagnostic procedure and surgical treatment, we targeted patients with more severe and actionable disease rather than all PA patients. Thus, based on the conditional independence assumption, our models suggest that amongst screened patients the prevalence of patients with similarly severe PA that are potentially candidates for surgery is approximately 7%. In contrast, while the method should construct models to rule-out phenotypes with a high negative predictive value, our selection of anchor variable is not ideal for this purpose because many patients with PA are not surgical candidates and not all have severe enough disease that they come to clinical attention. In addition, there are likely socioeconomic factors, care variability, and other logistic factors that are associated with anchor positivity. Properly adjusting for these effects will require more complex anchors and models. One natural extension of this strategy, for similarly complex phenotypes, would be derivation of a separate anchor variable and separate model for definitively negative PA controls.

Our models were estimated to have PPVs comparable to that of the originally evaluated rule-based algorithms and also identified several patients not flagged by these heuristics. But, as this was a proof-of-method concept, we focused on specific predictors selected by domain experts, did not exhaustively explore feature selection and engineering, and excluded patients with incomplete data from analysis. We would expect considerable further gains from more extensive modeling, and we plan to consider suitable missing data approaches to allow inclusion of all patients. In addition, our current method is suitable for developing phenotyping models when the number of potential predictors is far less than the number of records. It is of interest to explore additional predictors across high dimensional EHR data, which we expected would lead to models with improved accuracy and more precise estimates of anchor sensitivity. We plan to extend our current method to facilitate variable selection in building the prediction model.

## CONCLUSION

The incompleteness and asymmetry of EHR data limits its use for clinical decision support and research. We have developed a novel likelihood-based method that uses anchor-positive and unlabeled patients to simultaneously enable accurate model development, identification of unlabeled cases, and internal assessment of model performance. We expect this method will facilitate model development and transferability for a wide variety of EHR clinical decision support and research applications.

## CONTRIBUTORS

L.Z., D.S.H., and J.C. envisioned the study. L.Z. and J.C. developed the biostatistical methods. X.D. collected and processed the clinical data. I.A. and D.S.H. performed the chart review. L.Z., D.S.H., and J.C. wrote the manuscript. All authors approve of the final manuscript.

## FUNDING

This work was supported by University of Pennsylvania, Penn Medicine Precision Medicine Accelerator Grant, and NIH grants R56-HL138306 and R01-HL138306.

## COMPETING INTERESTS

None.

## ACKNOWLEDGEMENTS

We thank Debbie Cohen, Douglas Fraker, Scott Trerotola, and Heather Wachtel for creating and providing the list of patients registered in the primary aldosteronism research database.

**FIGURE LEGENDS**

**Figure 1.** Histogram of estimated probabilities of PA, $\hat{p}(Y = 1|X)$ for anchor 1 (A) and anchor 2 (B). The left panels display the probability histogram with the range of y-axis being 0-100%. The right panels display the zoomed-in probability histogram with y-axis limited to 0-6%. Anchor-positive (blue) and unlabeled patients (red) indicated.

# Supplementary Materials for "Electronic Health Record Phenotyping with Internally Assessable Performance (PhIAP) using Anchor-Positive and Unlabeled Patients"

Lingjiao Zhang, Xiruo Ding, Yanyuan Ma, Naveen Muthu, Imran Ajmal, Jason H. Moore, Daniel S. Herman, Jinbo Chen

**Appendix A: Proof of Parameter Identifiability**

Note that we have a random sample of $(\mathbf{X}, S)$'s, hence the observations $\mathbf{X}_i$'s form a random sample of $f(\mathbf{X})$. Thus, obviously, $f(\mathbf{X})$ is identifiable. Likewise, the observations $S_i$'s form a random sample of the pmf of $S$, hence $h$ is obviously identifiable. As derived above, the joint pdf of $(\mathbf{X}, S)$ is

$$f(\mathbf{X}) \left\{ \frac{hP(\mathbf{X};\boldsymbol{\beta})}{\int P(\mathbf{X};\boldsymbol{\beta})f(\mathbf{X})d\mathbf{X}} \right\}^{I(S=1)} \left\{ 1 - \frac{hP(\mathbf{X};\boldsymbol{\beta})}{\int P(\mathbf{X};\boldsymbol{\beta})f(\mathbf{X})d\mathbf{X}} \right\}^{I(S=0)}$$

with unknown parameters $h, \boldsymbol{\beta}, f(\mathbf{X})$. Note that the observed $(\mathbf{X}, S)$ is a random sample from the underlying population, thus $f(\mathbf{X})$ and $c$ are obviously identifiable. Assume that $\boldsymbol{\beta}$ is not identifiable. Then there exists $\boldsymbol{\beta}^*$ so that $p(\mathbf{X}, S; \boldsymbol{\beta}) = p(\mathbf{X}, S; \boldsymbol{\beta}^*)$ for all $(\mathbf{X}, S)$ pairs, which can be formalized as:

$$\left\{ \frac{hP(\boldsymbol{X};\boldsymbol{\beta})}{\int P(\boldsymbol{X};\boldsymbol{\beta})f(\boldsymbol{X})d\boldsymbol{X}} \right\}^{I(S=1)} \left\{ 1 - \frac{hP(\boldsymbol{X};\boldsymbol{\beta})}{\int P(\boldsymbol{X};\boldsymbol{\beta})f(\boldsymbol{X})d\boldsymbol{X}} \right\}^{I(S=0)}$$
$$= \left\{ \frac{hP(\boldsymbol{X};\boldsymbol{\beta}^*)}{\int P(\boldsymbol{X};\boldsymbol{\beta}^*)f(\boldsymbol{X})d\boldsymbol{X}} \right\}^{I(S=1)} \left\{ 1 - \frac{hP(\boldsymbol{X};\boldsymbol{\beta}^*)}{\int P(\boldsymbol{X};\boldsymbol{\beta}^*)f(\boldsymbol{X})d\boldsymbol{X}} \right\}^{I(S=0)}.$$

Letting $S = 1$, this equation reduces to

$$\frac{hP(\boldsymbol{X};\boldsymbol{\beta})}{\int P(\boldsymbol{X};\boldsymbol{\beta})f(\boldsymbol{X})d\boldsymbol{X}} = \frac{hP(\boldsymbol{X};\boldsymbol{\beta}^*)}{\int P(\boldsymbol{X};\boldsymbol{\beta}^*)f(\boldsymbol{X})d\boldsymbol{X}},$$

or equivalently

$$\frac{P(\boldsymbol{X};\boldsymbol{\beta}^*)}{P(\boldsymbol{X};\boldsymbol{\beta})} = \frac{\int P(\boldsymbol{X};\boldsymbol{\beta}^*)f(\boldsymbol{X})d\boldsymbol{X}}{\int P(\boldsymbol{X};\boldsymbol{\beta})f(\boldsymbol{X})d\boldsymbol{X}}. \tag{1}$$

Note that the right hand side of (1) is a positive constant, hence the left hand side is also a constant. When the covariates $X_j \to \infty$ if $\beta_j > 0$ and $X_j \to -\infty$ if $\beta_j \leq 0$, regardless what the intercept term is, $P(\boldsymbol{X};\boldsymbol{\beta}) \to 1$ and $P(\boldsymbol{X};\boldsymbol{\beta}^*)$ converges to either 1 or 0. Thus, the left hand side goes to either 1 or 0. Because the right hand side is positive, hence we further conclude that the left hand side goes to 1. Because it has to be a constant, this implies that the left hand side is always 1, i.e. $P(\boldsymbol{X};\boldsymbol{\beta}^*) = P(\boldsymbol{X};\boldsymbol{\beta})$ at all $\boldsymbol{X}$. Thus $\boldsymbol{\beta} = \boldsymbol{\beta}^*$, i.e. $\boldsymbol{\beta}$ is identifiable. Once $\boldsymbol{\beta}, h, f(\mathbf{X})$ are indentifiable, the disease prevalence $q = \int P(\boldsymbol{X};\boldsymbol{\beta})f(\boldsymbol{X})d\boldsymbol{X}$ is identifiable by definition.

**Appendix B: Phenotyping with an anchor of varying sensitivity**

Suppose that the population can be divided into $K$ strata of size $N_k$, where the $k^{th}$ stratum is indicated by $z_k$, $k = 1, ..., K$. Within each substratum, the anchor has a constant sensitivity. For notational convenience, $Z$ is included as a component of predictors $\boldsymbol{X}$. The problem is then formalized as stratified conditional independence,
$$p(S = 1|Y = 1, \boldsymbol{X}) = p(S = 1|Y = 1, Z = z_k) = c_k, k = 1, ..., K.$$

Model (**??**) remains the working model for prediction, where $Z$, if not predictive, has a log odds ratio equal to zero. The probability of observed data of an anchor-positive case and an unlabeled patient in each stratum is generalized as below to reflect variation in anchor sensitivity,

$$p(\boldsymbol{X}, S = 1) = c_k p(Y = 1|\boldsymbol{X}) p(\boldsymbol{X}),$$
$$p(\boldsymbol{X}, S = 0) = \{1 - c_k p(Y = 1|\boldsymbol{X})\} p(\boldsymbol{X}).$$

The likelihood function becomes the probability of observed data across all $K$ strata,

$$l(\boldsymbol{\beta}, \{c_k\}) \propto \sum_{i=1}^{N} \sum_{k=1}^{K} I(Z_i = z_k)[S_i \log\{c_k P(\boldsymbol{X}_i; \boldsymbol{\beta})\} + (1 - S_i) \log\{1 - c_k P(\boldsymbol{X}_i; \boldsymbol{\beta})\}]$$

Parameter identifiability can be shown similarly as in the situation of constant anchor sensitivity. The MLE of $\boldsymbol{\beta}$ and $\{c_k, \ k = 1, ..., K\}$ can be obtained by maximizing the log likelihood function $l(\boldsymbol{\beta}, \{c_k\})$. Let $q_k \equiv p(Y = 1|z_k)$ denote the phenotype prevalence in the $k^{th}$ stratum. Then $q_k$ can be estimated as $\hat{h}_k/\hat{c}_k$, where $\hat{h}_k$ is the estimated anchor prevalence in the $k^{th}$ stratum, $\hat{p}(S = 1|Z = z_k) = \sum_{i=1}^{N} I(Z_i = z_k, S_i = 1)/N_k$. The overall phenotype prevalence $q$ can then be estimated as a weighted summation of $q_k$'s, $\hat{q} = \sum_{k=1}^{K} N_k \hat{q}_k / N$.

**Appendix C: Predictive Accuracy Measures**

The accuracy measures are defined and further re-written as:

$$\begin{aligned}
TPR_v &= p\{P(\boldsymbol{X}; \boldsymbol{\beta}) > v | Y = 1, S = 0\} \\
&= p\{P(\boldsymbol{X}; \boldsymbol{\beta}) > v | Y = 1, S = 1\} \\
&= p\{P(\boldsymbol{X}; \boldsymbol{\beta}) > v | S = 1\} \\
FPR_v &= p\{P(\boldsymbol{X}; \boldsymbol{\beta}) > v | Y = 0, S = 0\} \\
&= \frac{p\{P(\boldsymbol{X}; \boldsymbol{\beta}) > v, S = 0\} - p\{P(\boldsymbol{X}; \boldsymbol{\beta}) > v, S = 0, Y = 1\}}{p(Y = 0, S = 0)} \\
&= \frac{p\{P(\boldsymbol{X}; \boldsymbol{\beta}) > v, S = 0\} - p\{P(\boldsymbol{X}; \boldsymbol{\beta}) > v | Y = 1, S = 0\} p\{Y = 1, S = 0\}}{p(Y = 0)} \\
&= \frac{p\{P(\boldsymbol{X}; \boldsymbol{\beta}) > v, S = 0\} - p\{P(\boldsymbol{X}; \boldsymbol{\beta}) > v | S = 1\} [p\{Y = 1\} - p\{Y = 1, S = 1\}]}{p(Y = 0)} \\
PPV_v &= p\{Y = 1 | P(\boldsymbol{X}; \boldsymbol{\beta}) > v, S = 0\} \\
&= \frac{p\{P(\boldsymbol{X}; \boldsymbol{\beta}) > v | Y = 1, S = 0\} p\{Y = 1, S = 0\}}{p\{P(\boldsymbol{X}; \boldsymbol{\beta}) > v, S = 0\}} \\
&= \frac{p\{P(\boldsymbol{X}; \boldsymbol{\beta}) > v | S = 1\} [p\{Y = 1\} - p\{Y = 1, S = 1\}]}{p\{P(\boldsymbol{X}; \boldsymbol{\beta}) > v, S = 0\}} \\
NPV_v &= p\{Y = 0 | P(\boldsymbol{X}; \boldsymbol{\beta}) < v, S = 0\} \\
&= \frac{p\{P(\boldsymbol{X}; \boldsymbol{\beta}) < v, S = 0\} - p\{P(\boldsymbol{X}; \boldsymbol{\beta}) < v, S = 0, Y = 1\}}{p\{P(\boldsymbol{X}; \boldsymbol{\beta}) < v, S = 0\}} \\
&= \frac{p\{P(\boldsymbol{X}; \boldsymbol{\beta}) < v, S = 0\} - p\{P(\boldsymbol{X}; \boldsymbol{\beta}) < v | S = 1\} pY = 1, S = 0}{p\{P(\boldsymbol{X}; \boldsymbol{\beta}) < v, S = 0\}} \\
AUC &= \int TPR_v dFPR_v
\end{aligned}$$

Thus they can be estimated as:

$$\widehat{TPR}_v = \frac{N^{-1}\sum_{i=1}^{N} I\{P(\boldsymbol{X}_i;\widehat{\boldsymbol{\beta}}) > v, S_i = 1\}}{N^{-1}\sum_{i=1}^{N} I(S_i = 1)}$$

$$\widehat{FPR}_v = \frac{N^{-1}\sum_{i=1}^{N} I\{P(\boldsymbol{X}_i;\widehat{\boldsymbol{\beta}}) > v, S_i = 0\}}{1 - N^{-1}\sum_{i=1}^{N} P(\boldsymbol{X}_i;\widehat{\boldsymbol{\beta}})}$$

$$- \frac{[N^{-1}\sum_{i=1}^{N} I\{P(\boldsymbol{X}_i;\widehat{\boldsymbol{\beta}}) > v, S_i = 1\}]\{N^{-1}\sum_{i=1}^{N} P(\boldsymbol{X}_i;\widehat{\boldsymbol{\beta}}) - N^{-1}\sum_{i=1}^{N} I(S_i = 1)\}}{\{N^{-1}\sum_{i=1}^{N} I(S_i = 1)\}\{1 - N^{-1}\sum_{i=1}^{N} P(\boldsymbol{X}_i;\widehat{\boldsymbol{\beta}})\}}$$

$$\widehat{PPV}_v = \frac{[N^{-1}\sum_{i=1}^{N} I\{P(\boldsymbol{X}_i;\widehat{\boldsymbol{\beta}}) > v, S_i = 1\}]\{N^{-1}\sum_{i=1}^{N} P(\boldsymbol{X}_i;\widehat{\boldsymbol{\beta}}) - N^{-1}\sum_{i=1}^{N} I(S_i = 1)\}}{\{N^{-1}\sum_{i=1}^{N} I(S_i =)\}[N^{-1}\sum_{i=1}^{N} I\{P(\boldsymbol{X}_i;\widehat{\boldsymbol{\beta}}) > v, S_i = 0\}]}$$

$$\widehat{NPV}_v = 1 - \frac{[N^{-1}\sum_{i=1}^{N} I\{P(\boldsymbol{X}_i;\widehat{\boldsymbol{\beta}}) < v, S_i = 1\}]\{N^{-1}\sum_{i=1}^{N} P(\boldsymbol{X}_i;\widehat{\boldsymbol{\beta}}) - N^{-1}\sum_{i=1}^{N} I(S_i = 1)\}}{\{N^{-1}\sum_{i=1}^{N} I(S_i =)\}[N^{-1}\sum_{i=1}^{N} I\{P(\boldsymbol{X}_i;\widehat{\boldsymbol{\beta}}) < v, S_i = 0\}]}$$

$$\widehat{AUC} = \int \widehat{TPR}_v d\widehat{FPR}_v.$$

**Appendix D: Simulation Studies**

We carried out extensive simulation studies to evaluate the performance of our method mainly in three aspects: (i) to demonstrate statistical consistency of our proposed estimators for $c$, $q$, and predictive accuracy measures; (ii) to evaluate the impact of model mis-specification on estimation of $c$ and $q$ and predictive accuracy; and (iii) to assess model performance when the anchor sensitivity is not fixed.

*Simulation study design* We generated the binary outcome variable $Y$ from the following logistic regression model:

$$\text{logit } p(Y=1|\boldsymbol{X};\boldsymbol{\beta}) = \beta_0 + \sum_{k=1}^{9} \beta_k X_k \quad (2)$$

with $(X_1, X_2, X_3)$, $(X_4, X_5, X_6)$ and $(X_7, X_8, X_9)$ representing groups of weak, moderate, and strong predictors, respectively. The three corresponding groups of parameter values were set as $(\beta_1, \beta_2, \beta_3) = (0.2, 0.4, 0.6)$, $(\beta_4, \beta_5, \beta_6) = (-1.0, -1.4, 1.8)$, and $(\beta_7, \beta_8, \beta_9) = (-2.0, 2.4, 2.8)$, respectively. We considered four different phenotype prevalences, $5\%, 10\%, 15\%$ and $20\%$, by setting the intercept parameter $\beta_0$ at values -2.5, 1.0, 3.3 and 5.4, respectively. The 9 predictors were independently distributed, with $X_1$, $X_4$ and $X_7$ generated from normal distribution $N(5, 10)$, $X_2$, $X_5$ and $X_8$ from Bernoulli distribution with success rate 0.5, and $X_3$, $X_6$ and $X_9$ from logit transformed standard uniform distribution. The anchor sensitivity $c = p(S = 1|Y = 1)$ was fixed at 0.5, so that $h$ takes four different values, 0.025, 0.05, 0.075 and 0.1, corresponding to the four different values of phenotype prevalence. In each Monte Carlo experiment, we first generated $\boldsymbol{X}$ for a target population with covariates as described above, then phenotype $Y$ according to model (2). For each case ($Y = 1$), the anchor variable $S$ was generated according to a Bernoulli distribution with parameter $c$. For the controls ($Y = 0$), we set $S = 0$. Then we drew a random sample of size $10,000$ as the training set, and pulled a disjoint testing set of size $5,000$. For each parameter combination, we iterated the simulation $1,000$ times. Below we focus our discussion on the results at phenotype prevalence $10\%$. Results for the other three prevalence values $5\%, 15\%$ and $20\%$ were similar and were included in the supplementary materials.

*Simulation results* When the data-generating model (2) above was fitted to each simulated dataset, the averaged parameter estimates ($\hat{\boldsymbol{\beta}}$) appeared very close to the true values ( Table S3). The negligibly small bias ($< 4\%$) indicated statistical consistency of the proposed MLE estimator. The averaged estimated asymptotic standard errors ("SE") and the empirical standard errors ("ESE") were very close with difference $< 6\%$. The anchor sensitivity $c$ was estimated to be 0.500 (SE: 0.021; 95% CI: 0.459, 0.541), and consequently the phenotype prevalence $q$ was estimated to be 0.100 (SE: 0.004; 95% CI: 0.094, 0.106) (Table S4). The fitted model was well calibrated as the nonparametrically-estimated and model-predicted cases numbers were comparable ( Table S5). PPV and TPR were estimated to be 0.798 (ESE: 0.056) and 0.852 (ESE: 0.024) respectively with decision threshold set at 0.5, which were very close to the "true values" (when $Y$ is truely observed for everyone) of PPV and TPR at 0.797 (ESE: 0.029) and 0.852 (ESE: 0.024). The AUC was estimated as 0.994 (ESE: 0.003), comparable to the "true value" of 0.993 (ESE: 0.001). The estimated predictive accuracy measures and their "true values" at other decision thresholds are summarized in Table S6.

To investigate the variation in model performance with respect to anchor sensitivity $c$, we considered two values for $c$, 0.5 and 0.2. With 10,000 observations in the training set and phenotype prevalence 10%, taking 0.5 as the decision threshold, the anchor variable helped identify 536 cases when $c$ equaled 0.5 and 797 cases when $c$ equaled 0.2. The estimates of $c$ and $q$ appeared to be consistent in both cases, although those at $c = 0.2$ had wider confidence intervals as expected (Table S4). The predictive accuracy of the fitted models were also comparable (Table S7). The OR estimates $\hat{\boldsymbol{\beta}}$ appeared somewhat biased when $c = 0.2$. However, the bias largely disappeared when the size of the training set was increased to 20,000 (Table S8). The model calibration also improved with the increased training set size (Table S9). The number of anchor-positive cases appeared to be most relevant in such bias reduction and calibration improvement in further unreported investigations.

We further explored the impact of model mis-specification on the performance of our proposed method. The models for analyzing EHR data are best considered as working models, and it is difficult to envision an underlying true model. By allowing deviation of the fitted model from the data generating model, we aimed to evaluate whether the resultant lack of fit is detectable by our method of assessing calibration and to evaluate whether our method of assessing calibration can inform unbiasedness of parameter estimation. Omission of weak predictors $(X_1, X_2, X_3)$ from model (2) had minor impact on model calibration and accuracy. However, omission of the strong $(X_7, X_8, X_9)$ predictors led to poor model calibration. The nonparametrically-estimated case numbers were very different from the model-predicted case numbers as shown in Table S5. Parameter estimation was sensitive to model misspecification as expected. The estimate of anchor sensitivity $c$ was 0.496 (SE: 0.023; 95% CI: 0.451, 0.541) and 0.358 (SE: 0.110; 95% CI: 0.142, 0.574) under the two mis-specified models respectively. The phenotype prevalence $q$ was correspondingly estimated as 0.101 (SE: 0.004; 95% CI: 0.093, 0.109) and 0.221 (SE: 0.036; 95% CI: 0.150, 0.292) respectively, where the bias in the latter estimate appeared to be biased as the 95% CI did not include true value 0.100 (Table S4). To summarize, the impact of model-misspecification on the performance of our method can be large, but can be detected through model calibration.

To assess the model performance when the anchor sensitivity varies, we simulated a new population following the same steps as above, except that it consisted of two strata indicated by a binary variable $Z$ that takes value 0 or 1 with probability $p(Z = 1) = 0.5$. Conditional on the phenotype the anchor variable was independent of all predictors within each strata, with $c_1 = p(S = 1|Y = 1, Z_1) = 0.2$, and $c_2 = p(S = 1|Y = 1, Z_2) = 0.8$. Applying the proposed method that takes the stratified conditional independence into account, the parameter estimates, $\hat{\boldsymbol{\beta}}$, $\hat{c}_1$, $\hat{c}_2$, $\hat{q}$, appeared consistent (Table S10). The anchor sensitivity $c_1$ was estimated to be 0.201 (SE: 0.025; 95% CI: 0.151, 0.251), $c_2$ was estimated to be 0.800 (SE: 0.022; 95% CI: 0.757, 0.843), and consequently the phenotype prevalence $q$ was estimated to be 0.100 (SE: 0.004; 95% CI: 0.091, 0.109). The fitted model was well calibrated and demonstrated good predictive accuracy (Table S11, S12). On the other hand, if the variation in anchor sensitivity failed to be recognized, the fitted model can lead to biased parameter estimates (Table S10), and notable decrease in predictive accuracy. The "true values" of $TPR$ dropped from 0.845 to 0.516 at decision threshold 0.5, and $AUC$ dropped from 0.993 to 0.976, although not much decrease was observed on $PPV$. (Table S12). To conclude, it is crucial to recognize the stratification, especially when the anchor sensitivities varies a lot.

**Appendix E: Supplementary Tables**

**Table S1.** Notes query table

| Key words | Expression |
|---|---|
| hyperaldo | hyperaldo |
| hyperaldo_spec | (?<!not\s)(?<!secondary\s)(?<!ruled out\s)(?<!ruling out\s)hyperaldo |
| primary_hyperaldo_spec | (?<!not\s)(?<!secondary\s)(?<!ruled out\s)(?<!ruling out\s)primary hyperaldo |
| primary_aldo | primary aldo |
| primary_aldo_spec | (?<!not\s)(?<!rule out\s)(?<!n't suggest\s)primary aldo |
| avs | adrenal vein sampl\|\bavs\b |
| bah | bilateral adrenal hyperplasia\|\bbah\b |
| bah_spec | (?<!not\s)bilateral adrenal hyperplasia\|(?<!not\s)\bbah\b |
| aldo_producing_adenoma | aldo\w{0,7}([-\s]producing)?([-\s]ade)?noma |
| aldo_producing_adenoma_spec | (?<!not\s\|nor\s)aldo\w{0,7}([-\s]producing)?([-\s]ade)?noma |
| adrenal_adenoma | adrenal adenoma |
| htn | (?<!pulmonary\s)(?<!pulmonary arterial\s)(htn\b\|hypertension) |
| htn_spec | (?<!pulmonary\s)(?<!pulmonary arterial\s)(?<!not\s\|nor\s)(htn\b\|hypertension) |
| htn_teixera | (?!pulm\w*\W*\w+\W+)\b(htn\|hypertension) |
| relative_hyperaldo | relative (hyper)?aldo |
| relative_hyperaldo_spec | (?<!not\s\|nor\s)relative (hyper)?aldo |
| salt_sensit | salt sensiti |
| salt_sensit_spec | (?<!not\s\|nor\s)(?<!not indicate )(?<!not consistent with hyperaldosteronism or )(?<!not show )salt sensiti |
| adrenalectomy | adrenalectomy |
| adrenalectomy_spec | (?<!not\s\|nor\s)adrenalectomy |
| word_count | \b\w+\b |

**Table S2.** Variable dictionary for the selected candidate predictors

| | Variable | Missing[1] | Description |
|---|---|---|---|
| Demographics | age | 0 | Age when aldosterone or renin test was performed (year) |
| | gender | 2 | Gender |
| | race | 177 | Race |
| | hisp | 69 | Hispanic (Yes/No) |
| Pre-visit | dbp | 546 | Diastolic blood pressure, from office visit closest (<= 14 days) to aldosterone/renin testing |
| | sbp | 546 | Systolic blood pressure, from office visit closest (<= 14 days) to aldosterone/renin testing |
| | time_bp_to_1st_RAR_yr | 2392 | Time interval (years) between first office visit with blood pressure recorded to aldosterone/renin test |
| | time_enc_to_1st_RAR_yr | 0 | Time interval (years) between first clinical encounter to aldosterone/renin test |
| Laboratory data | aldo | 0 | Serum aldosterone concentration (ng/dL) |
| | pra | 1168 | Plasma renin activity (ng/mL/hr) |
| | aldo:pra | 1168 | The aldosterone:renin ratio ((ng Aldosterone/dL)/(ng Angiotensin II/mL/hr)) |
| | test_potassium | 1520 | Blood potassium concentration (mmol/L) |
| | test_sodium | 1545 | Blood sodium concentration (mmol/L) |
| | test_carbon_dioxide | 1548 | Blood carbon dioxide concentration (mmol/L) |
| Encounter | enc_n | 0 | Number of clinical encounters |
| | enc_bp_n | 0 | Number of office visits with blood pressure recorded |
| | time_bp_after_1st_RAR_yr | 130 | Time interval (years) between aldosterone/renin test and last office visit with blood pressure |
| | time_enc_after_1st_RAR_yr | 0 | Time interval (years) between aldosterone/renin test and last clinical encounter |
| Diagnoses | Dx_h2_E26.0_9_n | 0 | Sum of the number of encounters with primary aldosteronism diagnosis codes (255.1, 255.10, 255.11, 255.12, E26.0, E26.01, E26.02, E26.09, E26.9) |
| | Dx_h2_E26.1_8_n | 0 | Sum of the number of encounters with other hyperaldosteronism diagnosis codes (255.13, 255.14, E26.1, E26.81, E26.89) |
| Notes | re_hyperaldo | 0 | count of 'hyperaldo' mentions in clinical notes |
| | re_primaryaldo | 0 | count of 'primary aldo' mentioned in the clinical notes |
| | re_bah | 0 | count of 'bah' mentioned in the clinical notes |
| | re_adrenal_adenoma | 0 | count of 'adrenal_adenoma' mentioned in the clinical notes |
| | re_htn | 0 | count of 'hypertension' mentioned in the clinical notes |
| | re_adrenalectomy | 0 | count of 'adrenalectomy' mentioned in the clinical notes |

[1]: Number of patients with missing data on each variable among the 6319 patients.

**Table S3.** Estimated odds ratio parameters when the phenotype prevalence was equal to 5%, 10%, 15%, or 20%.

| Phenotype prevalence | | $\beta_0$ | $\beta_1$ | $\beta_2$ | $\beta_3$ | $\beta_4$ | $\beta_5$ | $\beta_6$ | $\beta_7$ | $\beta_8$ | $\beta_9$ |
|---|---|---|---|---|---|---|---|---|---|---|---|
| *5%* | | | | | | | | | | | |
| | Mean | -2.629 | 0.208 | 0.433 | 0.634 | -1.053 | -1.472 | 1.899 | -2.112 | 2.545 | 2.952 |
| | Bias | -0.129 | 0.008 | 0.033 | 0.034 | -0.053 | -0.072 | 0.099 | -0.112 | 0.145 | 0.152 |
| | SE | 0.561 | 0.061 | 0.353 | 0.122 | 0.135 | 0.391 | 0.240 | 0.253 | 0.463 | 0.357 |
| | ESE | 0.573 | 0.062 | 0.368 | 0.127 | 0.144 | 0.416 | 0.258 | 0.269 | 0.476 | 0.383 |
| *10%* | | | | | | | | | | | |
| | Mean | 1.039 | 0.207 | 0.410 | 0.621 | -1.031 | -1.444 | 1.858 | -2.067 | 2.482 | 2.895 |
| | Bias | 0.039 | 0.007 | 0.010 | 0.021 | -0.031 | -0.044 | 0.058 | -0.067 | 0.082 | 0.095 |
| | SE | 0.471 | 0.046 | 0.267 | 0.092 | 0.101 | 0.294 | 0.181 | 0.189 | 0.345 | 0.268 |
| | ESE | 0.486 | 0.047 | 0.271 | 0.091 | 0.106 | 0.301 | 0.192 | 0.197 | 0.350 | 0.284 |
| *15%* | | | | | | | | | | | |
| | Mean | 3.401 | 0.202 | 0.406 | 0.616 | -1.023 | -1.441 | 1.842 | -2.049 | 2.461 | 2.868 |
| | Bias | 0.101 | 0.002 | 0.006 | 0.016 | -0.023 | -0.041 | 0.042 | -0.049 | 0.061 | 0.068 |
| | SE | 0.517 | 0.040 | 0.232 | 0.079 | 0.088 | 0.255 | 0.157 | 0.163 | 0.299 | 0.232 |
| | ESE | 0.520 | 0.041 | 0.237 | 0.082 | 0.091 | 0.258 | 0.166 | 0.171 | 0.308 | 0.237 |
| *20%* | | | | | | | | | | | |
| | Mean | 5.511 | 0.203 | 0.409 | 0.609 | -1.019 | -1.435 | 1.834 | -2.039 | 2.443 | 2.856 |
| | Bias | 0.111 | 0.003 | 0.009 | 0.009 | -0.019 | -0.035 | 0.034 | -0.039 | 0.043 | 0.056 |
| | SE | 0.583 | 0.036 | 0.209 | 0.072 | 0.079 | 0.231 | 0.143 | 0.148 | 0.270 | 0.211 |
| | ESE | 0.562 | 0.036 | 0.214 | 0.072 | 0.078 | 0.227 | 0.141 | 0.144 | 0.268 | 0.210 |

Mean: the mean $\hat{\beta}$ estimate; Bias: the difference between the mean $\hat{\beta}$ and the true value of $\beta$; SE: the mean asymptotic standard error estimate; ESE: the empirical standard error estimate

**Table S4.** Estimated $\hat{c}$ and $\hat{q}$ (95%CI) over 1000 replications. The logistic model was correctly specified (*"True Model"*), with three weak predictors (*"Misspecified − Model_1"*) omitted, or with three strong predictors (*"Misspecified − Model_2"*) omitted.

| | | *True Model* | *Misspecified − Model_1* | *Misspecified − Model_2* |
|---|---|---|---|---|
| $c = 0.5, q = 0.1$ | | | | |
| | $\hat{c}$ | 0.500 (0.459,0.541) | 0.496 (0.451,0.541) | 0.358 (0.142,0.574) |
| | $\hat{q}$ | 0.100 (0.094,0.106) | 0.101 (0.093,0.109) | 0.221 (0.150,0.292) |
| $c = 0.2, q = 0.1$ | | | | |
| | $\hat{c}$ | 0.200 (0.167,0.233) | 0.198 (0.163,0.233) | 0.131 (0.001,0.270) |
| | $\hat{q}$ | 0.101 (0.087,0.115) | 0.101 (0.087,0.115) | 0.364 (0.273,0.455) |

**Table S5.** Model-predicted vs. nonparametrically-estimated case numbers in the pre-defined intervals of predicted probabilities for 10% phenotype prevalence. The logistic model was correctly specified ("*True Model*"), with three weak predictors ("$Misspecified-Model_1$") omitted, or with three strong predictors ("$Misspecified-Model_2$") omitted. Results are presented as the mean estimate over 1000 iterations.

|          | $True Model$    |                              | $Misspecified-Model_1$ |                              | $Misspecified-Model_2$ |                              |
|----------|-----------------|------------------------------|-----------------|------------------------------|-----------------|------------------------------|
| Interval | Model predicted | Nonparametrically estimated  | Model predicted | Nonparametrically estimated  | Model predicted | Nonparametrically estimated  |
| 0.0_0.1  | 13              | 15                           | 18              | 19                           | 140             | 163                          |
| 0.1_0.2  | 13              | 14                           | 16              | 18                           | 197             | 234                          |
| 0.2_0.3  | 13              | 14                           | 17              | 18                           | 160             | 192                          |
| 0.3_0.4  | 14              | 15                           | 18              | 18                           | 122             | 149                          |
| 0.4_0.5  | 16              | 16                           | 20              | 20                           | 92              | 113                          |
| 0.5_0.6  | 19              | 18                           | 23              | 22                           | 68              | 88                           |
| 0.6_0.7  | 23              | 22                           | 28              | 26                           | 49              | 64                           |
| 0.7_0.8  | 30              | 29                           | 36              | 34                           | 35              | 49                           |
| 0.8_0.9  | 47              | 45                           | 54              | 52                           | 24              | 37                           |
| 0.9_1.0  | 313             | 312                          | 278             | 281                          | 296             | 161                          |

**Table S6.** Estimated vs. true predictive accuracy measures and their empirical standard errors of model performance when the phenotype prevalence was equal to $5\%, 10\%, 15\%$, or $20\%$. Results are presented as the mean estimate (the empirical standard error estimate).

| Phenotype prevalence | $cutoff$ | Estimated predictive accuracy[1] | | | | True predictive accuracy[2] | | | |
|---|---|---|---|---|---|---|---|---|---|
| | | 0.20 | 0.30 | 0.40 | 0.50 | 0.20 | 0.30 | 0.40 | 0.50 |
| *5%* | | | | | | | | | |
| | $\widehat{PPV}$ | 0.588 (0.056) | 0.658 (0.062) | 0.716 (0.069) | 0.767 (0.076) | 0.591 (0.036) | 0.662 (0.038) | 0.719 (0.041) | 0.770 (0.042) |
| | $\widehat{TPR}$ | 0.925 (0.02) | 0.892 (0.025) | 0.858 (0.03) | 0.822 (0.036) | 0.925 (0.02) | 0.892 (0.025) | 0.859 (0.03) | 0.822 (0.036) |
| | $\widehat{NPV}$ | 0.998 (0.001) | 0.997 (0.001) | 0.996 (0.001) | 0.995 (0.001) | 0.998 (0.001) | 0.997 (0.001) | 0.996 (0.001) | 0.995 (0.001) |
| | $\widehat{FPR}$ | 0.017 (0.003) | 0.012 (0.002) | 0.009 (0.002) | 0.007 (0.002) | 0.017 (0.002) | 0.012 (0.002) | 0.009 (0.002) | 0.007 (0.002) |
| | $\widehat{AUC}$ | 0.995 (0.003) | — | — | — | 0.995 (0.001) | — | — | — |
| *10%* | | | | | | | | | |
| | $\widehat{PPV}$ | 0.626 (0.043) | 0.694 (0.047) | 0.749 (0.051) | 0.798 (0.056) | 0.625 (0.026) | 0.693 (0.028) | 0.748 (0.028) | 0.797 (0.029) |
| | $\widehat{TPR}$ | 0.942 (0.012) | 0.914 (0.016) | 0.884 (0.02) | 0.852 (0.024) | 0.942 (0.012) | 0.914 (0.016) | 0.885 (0.02) | 0.852 (0.024) |
| | $\widehat{NPV}$ | 0.997 (0.001) | 0.995 (0.001) | 0.994 (0.001) | 0.992 (0.001) | 0.997 (0.001) | 0.995 (0.001) | 0.994 (0.001) | 0.992 (0.001) |
| | $\widehat{FPR}$ | 0.031 (0.004) | 0.022 (0.004) | 0.017 (0.004) | 0.012 (0.003) | 0.032 (0.003) | 0.023 (0.003) | 0.017 (0.003) | 0.012 (0.002) |
| | $\widehat{AUC}$ | 0.994 (0.003) | — | — | — | 0.993 (0.001) | — | — | — |
| *15%* | | | | | | | | | |
| | $\widehat{PPV}$ | 0.649 (0.035) | 0.714 (0.038) | 0.767 (0.041) | 0.812 (0.045) | 0.651 (0.021) | 0.717 (0.022) | 0.77 (0.023) | 0.815 (0.023) |
| | $\widehat{TPR}$ | 0.951 (0.009) | 0.926 (0.012) | 0.900 (0.015) | 0.870 (0.018) | 0.951 (0.009) | 0.927 (0.012) | 0.900 (0.015) | 0.870 (0.019) |
| | $\widehat{NPV}$ | 0.996 (0.001) | 0.993 (0.001) | 0.991 (0.001) | 0.989 (0.001) | 0.996 (0.001) | 0.993 (0.001) | 0.991 (0.001) | 0.989 (0.002) |
| | $\widehat{FPR}$ | 0.045 (0.005) | 0.032 (0.005) | 0.024 (0.004) | 0.018 (0.004) | 0.045 (0.004) | 0.032 (0.003) | 0.024 (0.003) | 0.017 (0.003) |
| | $\widehat{AUC}$ | 0.992 (0.004) | — | — | — | 0.992 (0.001) | — | — | — |
| *20%* | | | | | | | | | |
| | $\widehat{PPV}$ | 0.675 (0.03) | 0.737 (0.032) | 0.787 (0.034) | 0.829 (0.037) | 0.676 (0.019) | 0.738 (0.019) | 0.787 (0.019) | 0.83 (0.019) |
| | $\widehat{TPR}$ | 0.959 (0.007) | 0.938 (0.01) | 0.914 (0.012) | 0.887 (0.014) | 0.959 (0.007) | 0.937 (0.009) | 0.914 (0.012) | 0.887 (0.014) |
| | $\widehat{NPV}$ | 0.994 (0.001) | 0.992 (0.001) | 0.989 (0.001) | 0.985 (0.002) | 0.994 (0.001) | 0.992 (0.001) | 0.989 (0.002) | 0.985 (0.002) |
| | $\widehat{FPR}$ | 0.059 (0.006) | 0.043 (0.006) | 0.032 (0.005) | 0.023 (0.005) | 0.059 (0.005) | 0.043 (0.004) | 0.032 (0.004) | 0.023 (0.003) |
| | $\widehat{AUC}$ | 0.992 (0.005) | — | — | — | 0.991 (0.001) | — | — | — |

[1]: predictive accuracy measures according to our proposed method
[2]: predictive accuracy measures when $Y$ is truly observed

**Table S7.** Predictive accuracy measures of model performance at 10% phenotype prevalence when $c = 0.2$. Results are presented as the mean estimate (the empirical standard error estimate) over 1000 iterations.

|  | Estimated predictive accuracy[1] | | | | True predictive accuracy[2] | | | |
|---|---|---|---|---|---|---|---|---|
| $cutoff$ | 0.20 | 0.30 | 0.40 | 0.50 | 0.20 | 0.30 | 0.40 | 0.50 |
| $\widehat{PPV}$ | 0.728 | 0.779 | 0.818 | 0.851 | 0.727 | 0.779 | 0.819 | 0.852 |
|  | (0.05) | (0.051) | (0.054) | (0.058) | (0.032) | (0.034) | (0.035) | (0.036) |
| $\widehat{TPR}$ | 0.934 | 0.905 | 0.876 | 0.845 | 0.934 | 0.906 | 0.877 | 0.847 |
|  | (0.023) | (0.029) | (0.035) | (0.042) | (0.016) | (0.023) | (0.029) | (0.037) |
| $\widehat{NPV}$ | 0.994 | 0.992 | 0.989 | 0.986 | 0.994 | 0.992 | 0.989 | 0.986 |
|  | (0.002) | (0.002) | (0.003) | (0.003) | (0.001) | (0.002) | (0.003) | (0.003) |
| $\widehat{FPR}$ | 0.031 | 0.023 | 0.017 | 0.013 | 0.031 | 0.023 | 0.018 | 0.013 |
|  | (0.006) | (0.005) | (0.005) | (0.005) | (0.005) | (0.005) | (0.005) | (0.005) |
| $\widehat{AUC}$ | 0.994 | – | – | – | 0.993 | – | – | – |
|  | (0.006) | – | – | – | (0.001) | – | – | – |

[1]: predictive accuracy measures according to our proposed method
[2]: predictive accuracy measures when $Y$ is truly observed

**Table S8.** Estimated odds ratio parameters at 10% phenotype prevalence when $c = 0.2$.

|  | $\beta_0$ | $\beta_1$ | $\beta_2$ | $\beta_3$ | $\beta_4$ | $\beta_5$ | $\beta_6$ | $\beta_7$ | $\beta_8$ | $\beta_9$ |
|---|---|---|---|---|---|---|---|---|---|---|
| *Size of training set=10,000* | | | | | | | | | | |
| Mean | 1.192 | 0.220 | 0.453 | 0.664 | -1.115 | -1.573 | 2.029 | -2.243 | 2.698 | 3.143 |
| Bias | 0.192 | 0.020 | 0.053 | 0.064 | -0.115 | -0.173 | 0.229 | -0.243 | 0.298 | 0.343 |
| SE | 0.889 | 0.085 | 0.484 | 0.173 | 0.202 | 0.544 | 0.363 | 0.383 | 0.658 | 0.540 |
| ESE | 1.422 | 0.109 | 0.554 | 0.189 | 0.296 | 0.831 | 0.737 | 0.687 | 1.292 | 1.085 |
| *Size of training set=20,000* | | | | | | | | | | |
| Mean | 1.064 | 0.209 | 0.436 | 0.628 | -1.044 | -1.464 | 1.877 | -2.087 | 2.495 | 2.925 |
| Bias | 0.064 | 0.009 | 0.036 | 0.028 | -0.044 | -0.064 | 0.077 | -0.087 | 0.095 | 0.125 |
| SE | 0.577 | 0.055 | 0.320 | 0.111 | 0.123 | 0.354 | 0.220 | 0.230 | 0.416 | 0.327 |
| ESE | 0.594 | 0.056 | 0.334 | 0.117 | 0.125 | 0.351 | 0.226 | 0.237 | 0.403 | 0.332 |

Mean: the mean $\hat{\beta}$ over 1000 iterations; Bias: the difference between the mean $\hat{\beta}$ and the true value of $\beta$; SE: the mean asymptotic standard error estimate; ESE: the empirical standard error.

**Table S9.** Model-predicted vs. nonparametrically-estimated case numbers in the pre-defined intervals of predicted probabilities at 10% phenotype prevalence when $c = 0.2$. Results are presented as the mean estimate over 1000 iterations.

|  | *Size of training set=10,000* | | *Size of training set=20,000* | |
|---|---|---|---|---|
| Interval | Model predicted | Nonparametrically estimated | Model predicted | Nonparametrically estimated |
| 0.0_0.1 | 20 | 29 | 21 | 25 |
| 0.1_0.2 | 19 | 23 | 20 | 22 |
| 0.2_0.3 | 20 | 23 | 21 | 23 |
| 0.3_0.4 | 22 | 23 | 23 | 23 |
| 0.4_0.5 | 25 | 25 | 26 | 26 |
| 0.5_0.6 | 29 | 27 | 30 | 29 |
| 0.6_0.7 | 35 | 33 | 36 | 35 |
| 0.7_0.8 | 46 | 43 | 48 | 46 |
| 0.8_0.9 | 72 | 66 | 74 | 71 |
| 0.9_1.0 | 515 | 508 | 503 | 502 |

**Table S10.** Estimated ($\hat{\beta}$, $\hat{c}_1$, $\hat{c}_2$, $\hat{q}$) for stratified conditional independence with $c_1 = 0.2$ and $c_2 = 0.8$ at 10% phenotype prevalence.

| | $\beta_0$ | $\beta_1$ | $\beta_2$ | $\beta_3$ | $\beta_4$ | $\beta_5$ | $\beta_6$ | $\beta_7$ | $\beta_8$ | $\beta_9$ | $c_1$ | $c_2$ | $q$ |
|---|---|---|---|---|---|---|---|---|---|---|---|---|---|
| Varying anchor sensitivity[1] | | | | | | | | | | | | | |
| Mean | 1.046 | 0.207 | 0.410 | 0.622 | -1.031 | -1.437 | 1.856 | -2.062 | 2.478 | 2.882 | 0.201 | 0.800 | 0.100 |
| Bias | 0.046 | 0.007 | 0.010 | 0.022 | -0.031 | -0.037 | 0.056 | -0.062 | 0.078 | 0.082 | 0.001 | 0.000 | -0.000 |
| ASE | 0.559 | 0.041 | 0.238 | 0.081 | 0.089 | 0.262 | 0.159 | 0.165 | 0.459 | 0.234 | 0.025 | 0.022 | 0.004 |
| ESE | 0.577 | 0.040 | 0.243 | 0.082 | 0.091 | 0.258 | 0.161 | 0.168 | 0.476 | 0.237 | 0.025 | 0.022 | 0.004 |
| Constant anchor sensitivity[2] | | | | | | | | | | | | | |
| Mean | -2.338 | 0.148 | 0.293 | 0.445 | -0.735 | -1.025 | 1.324 | -1.471 | 4.827 | 2.047 | 0.712 | _ | 0.080 |
| Bias | -3.338 | -0.052 | -0.107 | -0.155 | 0.265 | 0.375 | -0.476 | 0.529 | 2.427 | -0.753 | 0.212 | _ | -0.021 |
| ASE | 0.363 | 0.033 | 0.195 | 0.064 | 0.065 | 0.210 | 0.116 | 0.118 | 0.327 | 0.168 | 0.028 | _ | 0.004 |
| ESE | 0.504 | 0.034 | 0.200 | 0.071 | 0.077 | 0.212 | 0.138 | 0.145 | 0.349 | 0.209 | 0.036 | _ | 0.005 |

Mean: the mean estimates over 1000 iterations; Bias: the difference between the mean estimates and the true values; SE: the mean asymptotic standard error estimate; ESE: the empirical standard error. [1]: the fitted model recognized the variation in anchor sensitivity; [2]: the fitted model failed to recognize the variation in anchor sensitivity.

**Table S11.** Model-predicted vs. nonparametrically-estimated case numbers in the pre-defined intervals of predicted probabilities for stratified conditional independence with $c_1 = 0.2$ and $c_2 = 0.8$ at 10% phenotype prevalence. Results are presented as the mean estimate over 1000 iterations.

| | Varying anchor sensitivity[1] | | Constant anchor sensitivity[2] | |
|---|---|---|---|---|
| Interval | Model predicted | Nonparametrically estimated | Model predicted | Nonparametrically estimated |
| 0.0_0.1 | 6 | 6 | 5 | 6 |
| 0.1_0.2 | 6 | 6 | 4 | 4 |
| 0.2_0.3 | 6 | 6 | 5 | 4 |
| 0.3_0.4 | 6 | 6 | 5 | 4 |
| 0.4_0.5 | 7 | 7 | 5 | 5 |
| 0.5_0.6 | 8 | 8 | 6 | 6 |
| 0.6_0.7 | 10 | 9 | 7 | 7 |
| 0.7_0.8 | 13 | 13 | 9 | 9 |
| 0.8_0.9 | 21 | 19 | 13 | 13 |
| 0.9_1.0 | 134 | 138 | 59 | 57 |

[1]: the fitted model recognized the variation in anchor sensitivity
[2]: the fitted model failed to recognize the variation in anchor sensitivity

**Table S12.** Estimated predictive accuracy measures of model performance for stratified conditional independence with $c_1 = 0.2$ and $c_2 = 0.8$ at 10% phenotype prevalence. Results are presented as the mean estimate (the empirical standard error).

|  | Varying anchor sensitivity[1] | | | | Constant anchor sensitivity[2] | | | |
|---|---|---|---|---|---|---|---|---|
| $cutoff$ | 0.20 | 0.30 | 0.40 | 0.50 | 0.20 | 0.30 | 0.40 | 0.50 |
| True predictive accuracy[3] | | | | | | | | |
| $\widehat{PPV}$ | 0.589 | 0.66 | 0.719 | 0.771 | 0.571 | 0.647 | 0.717 | 0.781 |
|  | (0.036) | (0.037) | (0.037) | (0.037) | (0.038) | (0.039) | (0.041) | (0.041) |
| $\widehat{TPR}$ | 0.938 | 0.908 | 0.879 | 0.845 | 0.684 | 0.618 | 0.565 | 0.516 |
|  | (0.022) | (0.028) | (0.033) | (0.038) | (0.065) | (0.066) | (0.066) | (0.066) |
| $\widehat{NPV}$ | 0.997 | 0.995 | 0.994 | 0.992 | 0.985 | 0.982 | 0.979 | 0.977 |
|  | (0.001) | (0.001) | (0.002) | (0.002) | (0.003) | (0.003) | (0.003) | (0.003) |
| $\widehat{FPR}$ | 0.032 | 0.023 | 0.017 | 0.012 | 0.025 | 0.016 | 0.011 | 0.007 |
|  | (0.004) | (0.004) | (0.003) | (0.003) | (0.003) | (0.003) | (0.002) | (0.002) |
| $\widehat{AUC}$ | 0.993 | 0.993 | 0.993 | 0.993 | 0.976 | 0.976 | 0.976 | 0.976 |
|  | (0.001) | (0.001) | (0.001) | (0.001) | (0.006) | (0.006) | (0.006) | (0.006) |
| Estimated predictive accuracy[4] | | | | | | | | |
| $\widehat{PPV}$ | 0.594 | 0.668 | 0.729 | 0.785 | 0.401 | 0.484 | 0.561 | 0.633 |
|  | (0.048) | (0.055) | (0.062) | (0.068) | (0.061) | (0.068) | (0.076) | (0.088) |
| $\widehat{TPR}$ | 0.944 | 0.918 | 0.889 | 0.858 | 0.910 | 0.873 | 0.835 | 0.790 |
|  | (0.015) | (0.018) | (0.021) | (0.024) | (0.021) | (0.026) | (0.03) | (0.036) |
| $\widehat{NPV}$ | 0.997 | 0.996 | 0.995 | 0.993 | 0.998 | 0.997 | 0.996 | 0.995 |
|  | (0.001) | (0.001) | (0.001) | (0.001) | (0.001) | (0.001) | (0.001) | (0.001) |
| $\widehat{FPR}$ | 0.031 | 0.022 | 0.016 | 0.012 | 0.034 | 0.023 | 0.016 | 0.011 |
|  | (0.005) | (0.004) | (0.004) | (0.004) | (0.004) | (0.004) | (0.003) | (0.003) |
| $\widehat{AUC}$ | 0.995 | — | — | — | 0.990 | — | — | — |
|  | (0.003) | — | — | — | (0.003) | — | — | — |

[1]: the fitted model recognized the variation in anchor sensitivity
[2]: the fitted model failed to recognize the variation in anchor sensitivity
[3]: predictive accuracy measures when $Y$ is truly observed
[4]: predictive accuracy measures according to our proposed method

**Table S13.** Estimated $\hat{c}$ (SE), $\hat{q}$ (SE) and $\widehat{AUC}$ for the series of models in model training - Anchor 1

|  | Covariates | $\hat{c}$ (SE) | $\hat{q}$ (SE) | $\widehat{AUC}$ |
|---|---|---|---|---|
| Demographics+pre_visit | age,gender,race,hisp,dbp,sbp, time_enc_to_1st_RAR,time_bp_to_1st_RAR | 0.104 (0.020) | 0.302 (0.054) | 0.787 |
| +Laboratory data | +aldo,aldo:pra,test_potassium, test_sodium,test_carbon_dioxide | 0.577 (0.119) | 0.047 (0.010) | 0.897 |
| +Encounters | +enc_n,enc_bp_n,time_enc_after_1st_RAR, time_bp_after_1st_RAR | 0.64 (0.131) | 0.054 (0.011) | 0.919 |
| +Diagnoses | +Dx_h2_E26.0_9_n,Dx_h2_E26.1_8_n | 0.486 (0.068) | 0.071 (0.010) | 0.963 |
| +Notes | +re_hyperaldo,re_primaryaldo,re_bah, re_adrenal_adenoma,re_htn,re_adrenalectomy | 0.451 (0.041) | 0.077 (0.007) | 0.994 |
| Final_model | age, aldo, aldo:pra, test_potassium, test_sodium, dbp,enc_bp,time_enc_after_1st_RAR, re_hyperaldo,re_adrenalectomy | 0.374 (0.047) | 0.076 (0.008) | 0.993 |

SE: Asymptotic standard error

**Table S14.** Estimated $\hat{c}$ (SE), $\hat{q}$ (SE) and $\widehat{AUC}$ for the series of models in model training - Anchor 2

|  | Covariates | $\hat{c}$ (SE) | $\hat{q}$ (SE) | $\widehat{AUC}$ |
|---|---|---|---|---|
| Demographics+pre_visit | age,gender,race,hisp,dbp,sbp, time_enc_to_1st_RAR,time_bp_to_1st_RAR | 0.148 (0.025) | 0.268 (0.041) | 0.780 |
| +Laboratory data | +aldo,aldo:pra,test_potassium, test_sodium,test_carbon_dioxide | 0.744 (0.107) | 0.049 (0.008) | 0.897 |
| +Encounters | +enc_n,enc_bp_n,time_enc_after_1st_RAR, time_bp_after_1st_RAR | 0.769 (0.114) | 0.058 (0.009) | 0.914 |
| +Diagnoses | +Dx_h2_E26.0_9_n,Dx_h2_E26.1_8_n | 0.541 (0.065) | 0.082 (0.010) | 0.972 |
| +Notes | +re_hyperaldo,re_primaryaldo,re_bah, re_adrenal_adenoma,re_htn,re_adrenalectomy | 0.562 (0.041) | 0.079 (0.006) | 0.993 |
| Final_model | aldo, aldo:pra, test_potassium, test_sodium, enc,enc_bp,time_bp_after_1st_RAR,Dx_h2_E26.0_9_n re_hyperaldo,re_primaryaldo,re_bah, re_adrenal_adenomare_adrenalectomy | 0.552 (0.042) | 0.070 (0.006) | 0.993 |

SE: Asymptotic standard error

**Table S15.** Final model - Anchor 1

|  | Intercept | age | aldo | aldo:pra | po | so | dbp | enc_bp | time_enc_after_1st_RAR_yr | re_hyperaldo | re_adrenalectomy |
|---|---|---|---|---|---|---|---|---|---|---|---|
| LogOR | -6.36 | -0.37 | 1.03 | 1.64 | 0.20 | 0.68 | -0.57 | -1.83 | 1.07 | 3.98 | 1.24 |
| SE | 0.69 | 0.30 | 0.40 | 0.37 | 0.27 | 0.27 | 0.28 | 0.39 | 0.27 | 1.00 | 0.36 |
| $p$ value | 0.00 | 0.22 | 0.01 | 0.00 | 0.46 | 0.01 | 0.04 | 0.00 | 0.00 | 0.00 | 0.00 |

SE: Asymptotic standard error

**Table S16.** Final model - Anchor 2

|  | Intercept | aldo | aldo:pra | po | so | enc | enc_bp | time_bp_after_1st_RAR_yr | hyperh260 | re_hyperaldo |
|---|---|---|---|---|---|---|---|---|---|---|
| LogOR | -6.46 | 0.99 | 1.10 | 0.20 | 0.25 | -0.11 | -2.12 | 0.98 | 1.36 | 2.01 |
| SE | 0.68 | 0.35 | 0.34 | 0.26 | 0.29 | 0.66 | 0.66 | 0.35 | 0.40 | 0.83 |
| $p$ value | 0.00 | 0.00 | 0.00 | 0.44 | 0.39 | 0.87 | 0.00 | 0.00 | 0.00 | 0.02 |

|  | re_primary_aldo | re_bah | re_adenoma | re_adrenalectomy |
|---|---|---|---|---|
| LogOR | 0.54 | 2.23 | 0.90 | 1.26 |
| SE | 0.28 | 0.83 | 0.27 | 0.37 |
| $p$ value | 0.06 | 0.01 | 0.00 | 0.00 |

SE: Asymptotic standard error

**Table S17.** Characteristics of patients

|  | Final population ($N = 3579$) | Rest ($N = 2740$) |
|---|---|---|
| *Age*[1] | 55 (16) | 54 (16) |
| *Gender* | | |
|   Male (%) | 1420 (40) | 1053 (38) |
|   Female (%) | 2159 (60) | 1687 (62) |
| *Race* | | |
|   Caucasian (%) | 1760 (50) | 1510 (57) |
|   African-American (%) | 1510 (43) | 930 (35) |
|   Other (%) | 234 (7) | 234 (8) |
| *Aldosterone*[2] (ng/DL) | 8.0 (4.5-14.9) | 8.0 (4.4-14.3) |
| *Renin*[2] (ng/mL/hr) | 0.9 (0.3-2.8) | 0.9 (0.3-2.6) |
| *Aldo* : *Renin*[2] | 8.0 (2.6-28.2) | 8.4 (2.9-26.4) |
| *Potassium*[1] (mmol/L) | 4.1 (0.6) | 4.1 (0.6) |
| *Sodium*[1] (mmol/L) | 139 (3.0) | 139 (3.5) |
| *CO2*[1] (mmol/L) | 26.7 (3.0) | 26.3 (4.1) |

[1]: Mean (SD), [2]: Median (IQR)

**Table S18.** Model-predicted vs. nonparametrically-estimated case numbers in the pre-defined intervals of predicted probabilities for the baseline models.

|  | Anchor 1 | | Anchor 2 | |
|---|---|---|---|---|
| Interval | *Model − predicted* | *Nonparametrically − estimated* | *Model − predicted* | *Nonparametrically − estimated* |
| 0.0_0.1 | 39 | 51 | 44 | 52 |
| 0.1_0.2 | 121 | 111 | 139 | 132 |
| 0.2_0.3 | 160 | 137 | 155 | 160 |
| 0.3_0.4 | 160 | 154 | 135 | 97 |
| 0.4_0.5 | 138 | 154 | 96 | 109 |
| 0.5_0.6 | 114 | 103 | 75 | 103 |
| 0.6_0.7 | 74 | 86 | 38 | 40 |
| 0.7_0.8 | 43 | 69 | 18 | 6 |
| 0.8_0.9 | 17 | 26 | 7 | 17 |
| 0.9_1.0 | 74 | 77 | 35 | 63 |